\documentclass[aps,prb,twocolumn,superscriptaddress,notitlepage]{revtex4-1}
\usepackage{graphicx}
\usepackage[caption=false]{subfig}
\usepackage{float}
\usepackage{natbib}
\usepackage{xcolor}
\usepackage{xr}
\usepackage{amsmath}
\usepackage{amssymb}
\usepackage{array}
\usepackage{wasysym}
\usepackage{color,soul}
\usepackage{braket}
\usepackage{verbatim}
\usepackage{hyperref}
\usepackage{gensymb}
\usepackage{url}
\usepackage{dirtytalk}

\usepackage{filecontents}
\begin{document}

\title{Experimental Evidence for the Spiral Spin Liquid in LiYbO$_2$}

\author{J. N. Graham}
\affiliation{School of Chemistry, University of Birmingham, Edgbaston, Birmingham B15 2TT, UK}
\affiliation{Institut Laue-Langevin, $71$ avenue des Martyrs, CS$20156$, $38042$ Grenoble C$\mathrm{\acute{e}}$dex $9$, France}
\author{N. Qureshi}
\affiliation{Institut Laue-Langevin, $71$ avenue des Martyrs, CS$20156$, $38042$ Grenoble C$\mathrm{\acute{e}}$dex $9$, France}
\author{C. Ritter}
\affiliation{Institut Laue-Langevin, $71$ avenue des Martyrs, CS$20156$, $38042$ Grenoble C$\mathrm{\acute{e}}$dex $9$, France}
\author{P. Manuel}
\affiliation{ISIS Neutron and Muon Source, Rutherford Appleton Laboratory, Harwell Campus, Didcot OX11 0QX, UK}
\author{A. R. Wildes}
\affiliation{Institut Laue-Langevin, $71$ avenue des Martyrs, CS$20156$, $38042$ Grenoble C$\mathrm{\acute{e}}$dex $9$, France}
\author{L. Clark}
\affiliation{School of Chemistry, University of Birmingham, Edgbaston, Birmingham B15 2TT, UK}

\date{\today}

\begin{abstract}
\noindent
Spiral spin liquids are an exotic class of correlated paramagnets with an enigmatic magnetic ground state composed of a degenerate manifold of fluctuating spin spirals. Experimental realisations of the spiral spin liquid are scarce, mainly due to the prominence of structural distortions in candidate materials that can trigger order-by-disorder transitions to more conventionally ordered magnetic ground states. Expanding the pool of candidate materials that may host a spiral spin liquid is therefore crucial to realising this novel magnetic ground state and understanding its robustness against perturbations that arise in real materials. Here, we show that the material LiYbO$_2$ is the first experimental realisation of a spiral spin liquid predicted to emerge from the $J_1$-$J_2$ Heisenberg model on an elongated diamond lattice. Through a complementary combination of high-resolution and diffuse neutron magnetic scattering studies on a polycrystalline sample, we demonstrate that LiYbO$_2$ fulfils the requirements for the experimental realisation of the spiral spin liquid and reconstruct single-crystal diffuse neutron magnetic scattering maps that reveal continuous spiral spin contours---a characteristic experimental hallmark of this exotic magnetic phase.

\end{abstract}
\maketitle
Magnetic states with unusual spin textures, such as magnetic skyrmions or vortices, are important both from a fundamental perspective as well as for their potential applications in emerging technologies. \cite{skyrmion1,skyrmion2,skyrmion3,skyrmion4} A lesser-explored manifestation of such unusual spin textures is the spiral spin liquid, a correlated paramagnetic state composed of a macroscopically degenerate manifold of fluctuating spin spirals that form closed contours or surfaces in reciprocal space. \cite{SSL_review,SpiralLiquids,SSL_1,SSL_2} From a theoretical perspective, spiral spin liquids have been predicted to exist in both two- and three-dimensions within the context of the frustrated $J_1$-$J_2$ Heisenberg model, for example, in honeycomb or diamond lattices. \cite{SSL_review,SSL_honeycomb,SSL_hyperhoneycomb} Experimentally, the search for the spiral spin liquid has focused on its realisation within cubic spinels, $AB_2X_4$, in which the $A$-site forms a three-dimensional diamond lattice. \cite{MnSc2S4_1,MnSc2S4_2,NiRh2O4_1,NiRh2O4_2,NiRh2O4_3,CoAl2O4,CoAl2O4_2,LiYbSe2, spinel_1,spinel_2,PaddisonMgCr2O4}

The diamond lattice can be considered as two interpenetrating FCC sublattices, with nearest-neighbour (NN), $J_1$, and next-nearest-neighbour (NNN), $J_2$, exchange interactions acting between and within sublattices, respectively (see Supplementary Information (SI), Fig. S1). \cite{SSL_1,SSL_2,diamond1} When the $J_1$ exchange interaction dominates, the diamond lattice is geometrically unfrustrated, leading to the formation of conventional ferromagnetic or N$\mathrm{\acute{e}}$el ordered magnetic ground states. However, the bipartite nature of the diamond lattice means that the exchanges can be readily tuned, and when $J_2/|J_1| \geq 1/8$, the system is expected to form a spiral spin liquid ground state.\cite{SSL_1,SSL_2,diamond1} Several cubic spinels have been identified as candidates for this model, including MnSc$_2$S$_4$ \cite{MnSc2S4_1,MnSc2S4_2}, NiRh$_2$O$_4$ \cite{NiRh2O4_1,NiRh2O4_2,NiRh2O4_3} and CoAl$_2$O$_4$. \cite{CoAl2O4,CoAl2O4_2} In particular, MnSc$_2$S$_4$ is considered the first experimental realisation of the spiral spin liquid on a perfect diamond lattice, with the observation of the characteristic spiral spin surface in single-crystal diffuse neutron magnetic diffraction measurements. \cite{MnSc2S4_1} However, experimentally, a true, continually fluctuating spiral spin liquid is still a rare occurrence. More often, the degeneracy of the system is lifted via an order-by-disorder mechanism, collapsing the fluctuating state into one with a single, long-range ordered spin helix. \cite{SSL_1,SSL_2,diamond1}

\begin{figure}
\centering
\includegraphics[width = 0.5 \textwidth]{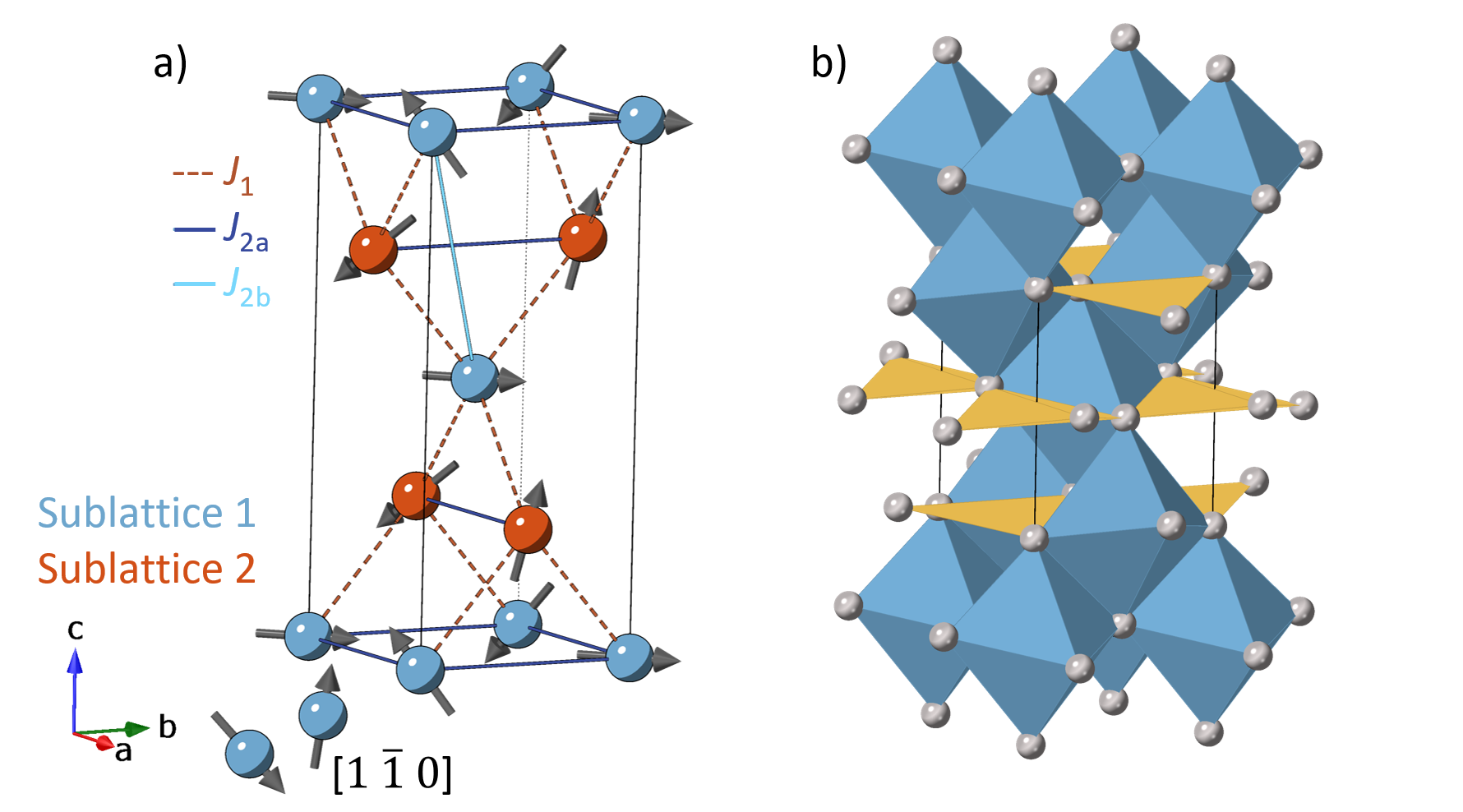}
\caption{a) The elongated diamond lattice---such as that formed from Yb$^{3+}$ moments in LiYbO$_2$---can be considered as two equivalent sublattices shown by the blue and orange spheres with NN $J_1$ (dashed orange) and NNN $J_2$ (solid blue) exchanges acting between and within sublattices, respectively. In the undistorted diamond lattice, $J_\mathrm{2a} = J_\mathrm{2b}$, but in the elongated diamond lattice model, $|J_1| \approx J_\mathrm{2a} > > J_\mathrm{2b}$, which introduces magnetic frustration. Moment orientations are shown by the grey arrows for the incommensurate helical structure with \textbf{k}$~= (0.3915(2), \pm 0.3915(2), 0)$ determined from WISH data analysis. Moments are shown propagating along the $[1~\bar{1}~0]$ direction. b) LiYbO$_2$ adopts the tetragonal $I4_1/amd$ structure where Yb$^{3+}$ ions (light blue polyhedra) form an elongated diamond network with tetrahedrally co-ordinated Li$^+$ ions (yellow polyhedra) lying in between. }
\label{Fig1}
\end{figure}

It was widely accepted that structural distortions to the perfect diamond lattice trigger this order-by-disorder mechanism, and therefore, are detrimental to the realisation of the spiral spin liquid phase. However, a recent re-imagining of this theory determined that a spiral spin liquid could still be realised under a tetragonal distortion to the diamond lattice. \cite{SSL_1,diamond1,Bordelon} The key to this elongated diamond lattice model is to consider the effect of the tetragonal distortion on the superexchange pathways. In the perfect diamond lattice, $J_2$ describes the next-nearest neighbour superexchange pathway, but in the elongated model, $J_2$ is spilt into two nonequivalent exchanges, $J_{2\mathsf{a}}$ and $J_{2\mathsf{b}}$ (Fig. \ref{Fig1}a). Now, the superexchange pathways $J_1$ and $J_{2\mathsf{a}}$ ($J_2\parallel\langle100\rangle$ in the undistorted lattice) may be nearly equivalent in length and substantially shorter than the $J_{2\mathsf{b}}$ pathway ($J_2\parallel\langle111\rangle$ in the undistorted lattice). In this case, the strength of the $J_{2\mathsf{b}}$ exchange interaction is assumed negligible in comparison to $J_1$ and $J_{2\mathsf{a}}$. If the magnitudes of $J_1$ and $J_{2a}$ are comparable, this elongated diamond lattice is expected to be frustrated and may lead to the emergence of a novel spiral spin liquid ground state. 

One potential candidate for the realisation of this $J_1$-$J_2$ Heisenberg model on an elongated diamond lattice is NaCeO$_2$. This system adopts a tetragonal $I4_1/amd$ structure with an axial elongation of $c/\sqrt{2}a = 1.63$, which was considered sufficient to render $J_{2b}$ negligible. \cite{NaCeO2} However, neutron diffraction data for NaCeO$_2$ show that it forms a commensurate N$\mathrm{\acute{e}}$el ordered ground state below $3.18~$K due to $J_1 >> J_{2a}$. This is not surprising, given theory predicts that when $J_{2a}>0$ (antiferromagnetic), the spiral spin liquid phase exists within a relatively narrow region between commensurate ferromagnetic ($-4 < J_{1}/J_{2a}$) and antiferromagnetic ($J_1/J_{2a} < 4$) phases. \cite{diamond1,Bordelon} Thus, here we turn our attention to LiYbO$_2$, which like NaCeO$_2$ adopts a tetragonal $I4_1/amd$ structure  (Fig. \ref{Fig1}b), but importantly, the ratio of exchange parameters has been determined as $J_1 = 1.426J_{2a} > 0$, placing LiYbO$_2$ within the boundary of the spiral spin liquid phase of the elongated diamond lattice. \cite{Bordelon,LiRO2} 

However, a recent investigation of LiYbO$_2$ concluded that it is not an experimental realisation of the spiral spin liquid. \cite{Bordelon,Kenney} Theory allows for an incommensurate helical structure propagating along the diagonal of the tetragonal basal plane (\textbf{k}$~= (\delta, \pm \delta, 0)$), which acts across the two Yb$^{3+}$ sublattices. Critically, however, the moments on these sublattices must have a fixed phase difference of $\phi = \pi$ to allow moments to rotate along all bonds between the sublattices equivalently. \cite{Bordelon} Experimentally, $\phi$ was found to be only $0.58~\pi$ for LiYbO$_2$, resulting in a staggering effect between moments, and additional perturbations to the $J_1$-$J_2$ Heisenberg model on the elongated diamond lattice were required to explain these previous experimental results. The possibility for disorder, for example, in the local Yb$^{3+}$ environment, was proposed as the cause for this discrepancy in the phase angle, but structural disorder was not resolvable within the diffraction measurements presented. \cite{Bordelon} Furthermore, the ordered moment size determined from previous powder neutron diffraction measurements was only $84\%$ of the expected full ordered moment of the $S_{\mathsf{eff}}={\frac{1}{2}}$ Yb$^{3+}$ ions in LiYbO$_2$, indicating that the true nature of its ground state remained elusive. Thus, here we combine complementary high-resolution and diffuse neutron magnetic scattering measurements on LiYbO$_2$ to recover the full Yb$^{3+}$ moment and re-examine the phase angle, $\phi$. In doing so, we identify LiYbO$_2$ as the first experimental realisation of the $J_1$-$J_2$ Heisenberg spiral spin liquid on the elongated diamond lattice.

\begin{figure}
\centering
\includegraphics[]{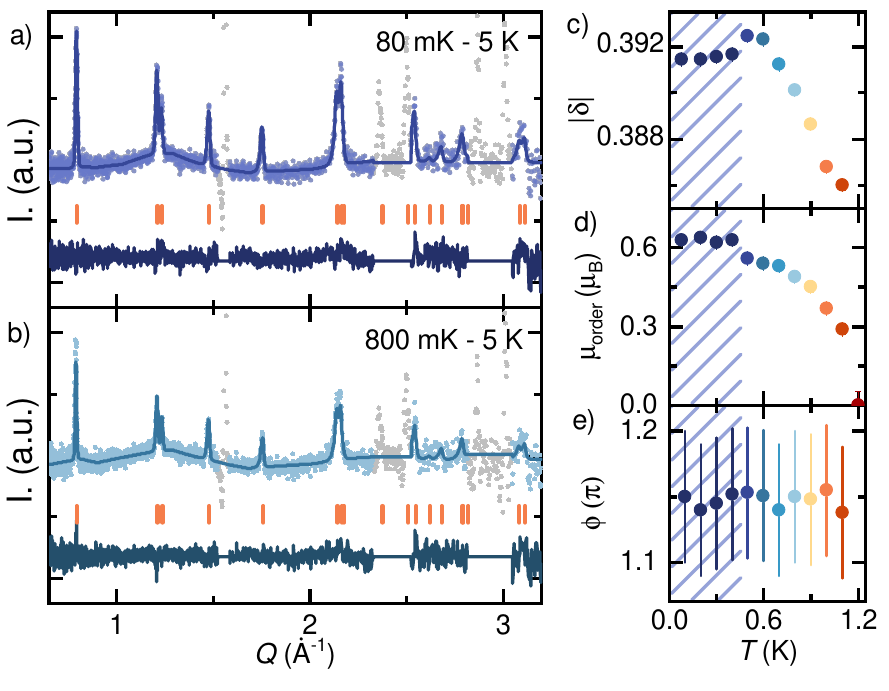}
\caption{Refinement of the incommensurate helical magnetic structure within the spiral spin liquid phase of LiYbO$_2$ at a) $80~$mK ($R_\mathrm{mag} = 13.1~\%, \chi^2 = 3.06$ and b) $800~$mK ($R_\mathrm{mag} = 7.49~\%, \chi^2 = 2.96$) from data collected on WISH. In both a) and b) data points and fits are shown in blue, difference curves in dark blue, magnetic Bragg reflections by the orange tickmarks and excluded data from the paramagnetic subtraction in grey. Evolution of refined parameters c) propagation vector, \textbf{k} = $(\delta, \pm \delta, 0)$, d) ordered moment size, $\mu_\mathrm{order}$ and e) phase angle, $\phi$. A transition occurs at $450~$mK, where \textbf{k} and $\mu_\mathrm{order}$ change from evolving as a function of temperature to being constant in temperature. This locking-in of the parameters coincides with a peak in previous heat capacity measurements of LiYbO$_2$.\cite{Bordelon}}
\label{Fig2}
\end{figure}

A high-quality polycrystalline sample of LiYbO$_2$ was prepared via a solid-state synthesis method described in the SI. Time-of-flight neutron powder diffraction data were collected on the long-wavelength WISH diffractometer at the ISIS Neutron and Muon Source. \cite{WISH, WISH_exp_LYO} Magnetic Bragg peaks measured between $0.08~$K to $1.2~$K were isolated by subtracting a paramagnetic dataset collected at $5~$K. Below $450~$mK there are several magnetic Bragg peaks in the temperature-subtracted diffraction data that may be indexed by an incommensurate helical magnetic model. Fig.\ref{Fig2}a shows the Rietveld refinement of this model against the data measured at $80~$mK, where the propagation vector was refined to the doubly degenerate \textbf{k}~= (0.3915(2), $\pm0.3915(2),~0)$ with a schematic of the corresponding magnetic structure shown in Fig. \ref{Fig1}a. \cite{Fullprof,Mag2Pol} The refined moment, $\mu_{\mathrm{order}} = 0.63(1)~\mu_\mathrm{B}$ is only $\sim40\%$ of the expected full ordered moment size, $\mu_{\mathrm{exp}} = gS \mu_{\mathrm{B}} = 1.5~\mu_{\mathrm{B}}$, assuming $S_{\mathsf{eff}} = 1/2$ and $g = 3$.\cite{Bordelon} The phase angle between the two sublattices of Yb$^{3+}$ moments refined to $\phi = 1.15(5)~\pi$. Refinements as a function of temperature have no variation below $450$~mK in \textbf{k}, $\mu_{\mathsf{order}}$ or $\phi$ (Figs. {\ref{Fig2}c, d and e}, respectively).
\color{black}
\\
\indent
Previous characterisation of LiYbO$_2$ discounted its candidacy as a spiral spin liquid based on the measured phase angle between the sublattices, $\phi = 0.58~\pi$, being in disagreement with the theoretical prediction of $\phi = \pi$. \cite{Bordelon} However, the refinement of the phase angle in this work, $\phi = 1.15(5)~\pi$ is consistent with the theoretical prediction and, therefore, verifies that LiYbO$_2$ maps onto the $J_1$-$J_2$ Heisenberg model on an elongated diamond lattice. In the previous experimental study, this misassignment of $\phi$ likely stemmed from the need to pelletise the polycrystalline sample for the in-field neutron diffraction measurements presented. This in turn required the inclusion of a significant preferred orientation correction during the Rietveld analysis of the powder neutron diffraction data. \cite{Bordelon} All measurements presented here were performed on a loose powder, therefore, preferred orientation is not an issue, as confirmed through Rietveld refinement of the chemical structure against high-resolution powder neutron difffraction data collected on the D2B instrument at the Institut Laue-Langevin (ILL) using \texttt{Mag2Pol} (SI). \cite{D2Bpaper,D7Data, Mag2Pol} Additionally, the magnetic structure is best described by a propagation vector of \textbf{k}$~=~(\delta, \pm \delta, 0)$, which is the correct form required to generate the spiral spin liquid phase. \cite{Bordelon,diamond1,PhysRevLett.101.047201} The ratio of exchange parameters determined from this helical structure, $J_1 = 1.343(4)~J_{2a~}>~0$ places LiYbO$_2$ directly within the spiral spin liquid phase. Therefore, we propose that LiYbO$_2$ is the first material to fulfil experimentally all of the conditions required for the observation of the spiral spin liquid in the $J_1$-$J_2$ Heisenberg model on an elongated diamond lattice. \cite{Bordelon} 

In addition to this spiral spin liquid ground state, time-of-flight neutron powder diffraction data reveal that an intermediary phase exists between $450~$mK and $1.2~$K. This phase is similar to the one determined below $450~$mK but evolves with temperature. An example of the Rietveld analysis of this intermediary phase is shown in Fig. \ref{Fig2}b for a dataset measured on WISH at $800~$mK. Refinement as a function of temperature reveals three key conclusions about the intermediary phase: (1) the propagation vector, \textbf{k}, varies with temperature but always takes the form \textbf{k} = $(\delta, \pm \delta, 0)$, (2) the ordered moment size, $\mu_\mathrm{order}$, gradually reduces from $0.63(1)~\mu_\mathrm{B}$ to zero by $1.2~$K, but (3) the phase angle between the sublattices, $\phi = 1.15(5)~\pi$ remains constant throughout the phase, as shown in Figs. \ref{Fig2}c, d and e, respectively. Therefore, this intermediary phase still fulfils the requirements of the spiral spin liquid.\cite{Bordelon}

Whilst from an average-structure perspective the spiral spin liquid phase in LiYbO$_2$ has now been confirmed, a remaining question is to understand why the experimentally determined moment size is significantly reduced in comparison to the full ordered moment. The defining experimental signature of the spiral spin liquid is a broad continuous ring of neutron magnetic scattering in reciprocal space \cite{MnSc2S4_1}, caused by the correlated fluctuations of spin spirals. Thus, it seems plausible that the remaining missing moment must be contained within these spin spiral fluctuations, which will give the characteristic diffuse magnetic scattering of the spiral spin liquid. To confirm this hypothesis, diffuse polarised neutron scattering measurements were performed on D7 at the ILL \cite{D7, Schweika_2010, D7Data} as described in the SI. 
\begin{figure}
\centering
\includegraphics[]{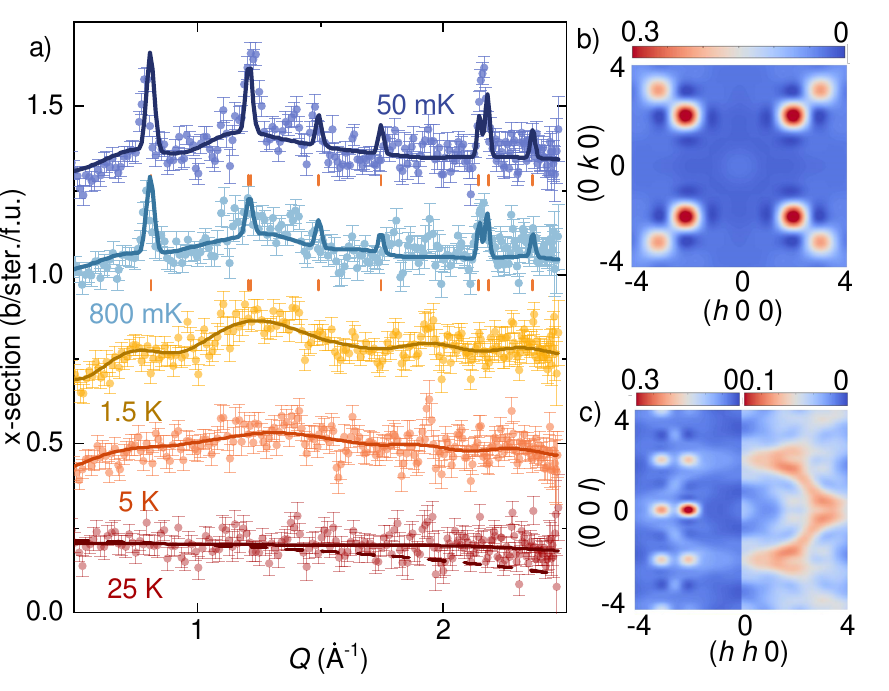}
\caption{a) Neutron magnetic scattering cross-section for LiYbO$_2$ as measured on D7. At $50~$mK and $800~$mK, both magnetic Bragg and correlated diffuse scattering are observed for the spiral spin liquid phase, which were analysed simultaneously with \texttt{SPINVERT + Bragg} by making the commensurate approximation to \textbf{k}$~=~(0.4, \pm 0.4, 0)$ (orange tickmarks). Data at $1.5~$K and $5~$K were fit with \texttt{SPINVERT} (solid lines) and $25~$K by a paramagnetic form factor of Yb$^{3+}$ ions (dashed line). Data have been vertically shifted by $0.3~$b./ster./f.u. for clarity. Single-crystal reconstructions from powder data analysis at $50~$mK, b) positions of incommensurate magnetic Bragg peaks in the $(h~k~0)$ plane and c) the signature of the spiral-spin liquid, a continuous diffuse ring in the $(h~h~l)$ plane (left: total magnetic scattering, right: magnetic diffuse scattering only, magnetic Bragg peaks have been excluded to show the diffuse ring more clearly). Intensities have been normalised to absolute units according to the colour bars. }
\label{Fig3}
\end{figure}
The time-equal magnetic cross-sections of LiYbO$_2$ measured on D7 are presented in Fig.\ref{Fig3}a. Paramagnetic behaviour is observed in the magnetic scattering at $25~$K and was modelled using a $\langle j_0\rangle$ analytical approximation of the Yb$^{3+}$ form factor \cite{Magff}. The approximation assumes a $S_\mathrm{eff} = 1/2$ ground state for Yb$^{3+}$ with $g = 3$, which is valid given the integration window of D7 ($-20 \leq \Delta E < 3~$meV) is well below the previously observed crystal field excitations for LiYbO$_2$. \cite{Bordelon} Beneath the magnetic Bragg peaks at $50~$mK and $800~$mK there is significant structure in the diffuse magnetic scattering, which is also visible in the WISH data (Fig. \ref{Fig2}). Most prominent is the broad feature centred around $Q = 1.2~$\r{A}$^{-1}$ which has virtually no change in size or shape in the diffuse scattering between $50~$mK and $1.5~$K. 

The $50~$mK and $800~$mK data were modelled using a modified version of the \texttt{SPINVERT} program (\texttt{SPINVERT + Bragg}) in order to simultaneously fit the magnetic Bragg and diffuse features (SI \cite{helix, FENNELL201724}). \cite{Spinvert1,Spinvertbragg} A challenging aspect of this analysis was the incommensurate nature of the magnetic structure, as \texttt{SPINVERT} relies on periodic boundary conditions to generate a supercell. Therefore, the propagation vector determined on WISH was approximated to the commensurate \textbf{k} = (0.4, $\pm0.4$, 0), and a supercell of $10 \times 10 \times 4$ unit cells was constructed. Spins were allowed complete rotational freedom, as introducing anisotropy to refine in a conical fashion did not make any difference to the fit. The resultant fits are shown in Fig.\ref{Fig3}a and emulate the main features of the data extremely well. Additionally, the diffuse-only data measured at $1.5~$K and $5~$K were fit with \texttt{SPINVERT} using a supercell of $6 \times 6 \times 3$ unit cells. The data from D7 are normalised to an absolute intensity scale, and therefore, the scale factor from \texttt{SPINVERT} is directly related to the moment size. This was separated into the proportion of the moment which is ordered and disordered, and these values are summarised in Table \ref{Moments}. 
\noindent
\begin{table}
\begin{tabular}{c|c|ccc}
\hline
\hline
$T$ (K) & WISH & &D7 &\\
 & $\mu\mathrm{_{order}}$ ($\mu\mathrm{_B}$) & $\mu\mathrm{_{order}}$ ($\mu\mathrm{_B}$) &  $\mu\mathrm{_{disorder}}$ ($\mu\mathrm{_B}$) & $\mu_{\mathrm{total}}$   ($\mu\mathrm{_B}$)\\
\hline
Base & 0.63(1) & 0.80(1) & 1.80(1) & 2.60(1)\\
0.8 & 0.49(1) & 0.67(2) & 1.92(2) & 2.59(2)\\
1.5 & - & - & 2.73(2) & 2.73(2)\\
5 & - & - & 2.74(2) & 2.74(2)\\
25 & - & - &2.76(2) & 2.76(2)\\
\hline
\hline
\end{tabular}
\caption{Summary of experimental magnetic moments for LiYbO$_2$, with $\mu_{\mathrm{order}}$ from Rietveld analysis of WISH data and $\mu_{\mathrm{total}}$ split into helically ordered, $\mu_{\mathrm{order}}$ and correlated short-range ordered, $\mu_{\mathrm{disorder}}$ components from \texttt{SPINVERT + Bragg} analysis of D7 data. Base temperatures for the WISH and D7 experiments are $80~$mK and $50~$mK, respectively.}
\label{Moments}
\end{table}

Thus, we show that the full moment of LiYbO$_2$ can be obtained through the simultaneous refinement of magnetic Bragg and diffuse scattering. It is important to note that the moments obtained from each instrument are not directly comparable due to the definitions used, Rietveld method: $\mu^2 = g^2 S^2~\mu_\mathrm{B}^2$ and \texttt{SPINVERT + Bragg} analysis: $\mu^2=g^2 S(S+1)~\mu_\mathrm{B}^2$. Assuming both have $S_{\mathsf{eff}} = 1/2$ and $g =3$, \cite{Bordelon} a full moment size of $\mu= 1.5~\mu_{\mathrm{B}}$ for WISH and $\mu = 2.6~\mu_{\mathrm{B}}$ for D7 is expected. At base temperature, $\mu_{\mathrm{order}}$ accounts for $40\%$ and $30\%$ of the expected moments for WISH and D7, respectively. The difference between the two methods is likely due to the incommensurate-commensurate approximation in the \texttt{SPINVERT + Bragg} analysis which underfits some of the magnetic Bragg peaks, most prominently, at $Q~=~2.16~$\r{A}$^{-1}$. This is a limitation of the reverse Monte Carlo analysis at this time, however, given the complexity of the magnetic structure, and the agreement with the WISH results, we conclude that the moment sizes presented here are an accurate description of the system. 

Reconstructions of single-crystal neutron magnetic diffraction patterns were calculated from fits of the $50~$mK D7 powder data using \texttt{SCATTY}. \cite{Spinvert1,Spinvert2} Figure \ref{Fig3}b shows the simulation for the $(h~k~0)$ plane, which reveals strong magnetic intensity at the expected magnetic Bragg peak positions. Due to the calculation method of the single-crystal scattering plane, the magnetic Bragg peaks appear artificially broadened in the simulation. The ($h~h~l$) plane, being a slice diagonally through the ($h~k~0)$ plane, is shown in Fig.\ref{Fig3}c, where the ring of diffuse scattering, representing the contoured spiral surface, can clearly be seen. This is the expected manifestation of single-crystal  diffuse magnetic neutron scattering for a spiral spin liquid \cite{MnSc2S4_1} and, therefore, further corroborates that the analysis and conclusions presented here for the magnetic ground state of LiYbO$_2$ are correct.

The spin-spin correlation function, $\langle$\textbf{S}$(0)\cdot$\textbf{S}$(r)\rangle$, for LiYbO$_2$ was calculated by \texttt{SPINCORREL} \cite{Spinvert1,Spinvert2} from fits to the D7 data. 
\begin{figure}
\centering
\includegraphics[]{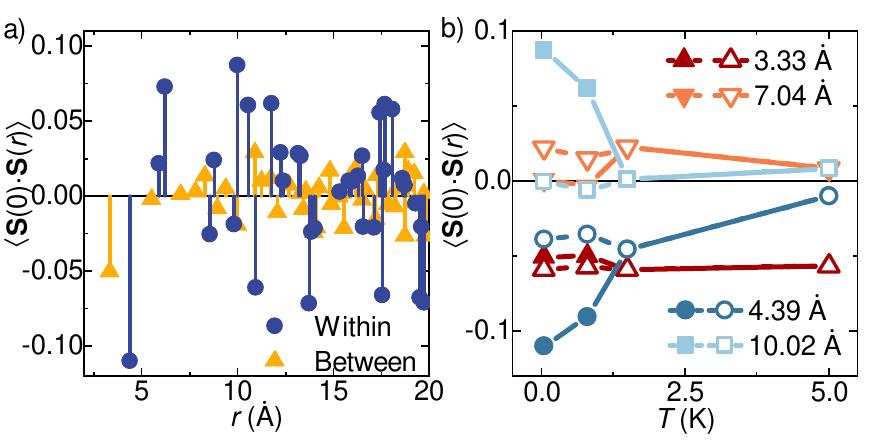}
\caption{a) Spin-spin correlation function, $\langle$\textbf{S}$(0)\cdot$\textbf{S}$(r)\rangle$, as a function of interatomic distance, $r$, for LiYbO$_2$ at $50~$mK. Correlations are divided into those within (blue) and between sublattices (yellow). b) Spin-spin correlation function,  $\langle$\textbf{S}$(0)\cdot$\textbf{S}$(r)\rangle$, as a function of temperature showing the relative contributions of the short-range (open, constant) and long-range (closed, decreasing) correlations.}
\label{Fig4}
\end{figure}
The correlations at $50~$mK (Fig. \ref{Fig4}a) are in good agreement with the predicted correlations for a helically ordered state (SI). The modified \texttt{SPINVERT + Bragg} analysis allows for the proportion of correlations to be broken down into short- and long-range ordered contributions, as shown by the open and closed markers, respectively in Fig. \ref{Fig4}b. The temperature dependence of these correlations quantifies the observations made in the D7 data. Namely, correlations that contribute to the diffuse scattering are temperature-independent between $50~$mK and $1.5~$K but ordered contributions are temperature-dependent. 

Taken together, this complementary neutron scattering study shows that a new family of materials---those with an elongated diamond lattice---provides an experimentally achievable route to the spiral spin liquid. There are three conditions needed to confirm the spiral spin liquid phase: (1) a phase angle, $\phi$, between the magnetic sublattices that is equal to $\pi$, (2) a propagation vector of the form \textbf{k}~$ = (\mathrm{q}, \mathrm{q}, 0)$ and (3) a ratio of exchange interactions of $-4 < J_1/|J_2| < 4$. Through the results presented here, we have confirmed conditions (1) $\phi = 1.15(5)~\pi$, (2) \textbf{k}$~= (\delta, \pm \delta, 0)$ and (3) $J_1 = 1.343(4)~J_2a~>~0$ below $450~$mK. Further, conclusive evidence for the spiral spin liquid phase was revealed by the continuous spin spiral contours in the $(h~h~l)$ plane from single-crystal reconstructions of the diffuse powder scattering data. Future studies on single crystals would thus confirm the exchange parameters through inelastic neutron scattering and verify our magnetic diffuse scattering simulations. The expected full moment of LiYbO$_2$ was recovered through the analysis of diffuse neutron scattering measurements and is split between the long- and short-range contributions characteristic of the spiral spin liquid. To date, only LiYbO$_2$ \cite{Bordelon,Kenney} and NaCeO$_2$ \cite{NaCeO2} have been explored within the context of the $J_1$-$J_2$ Heisenberg model on the elongated diamond lattice, but their differing ratios of exchange parameters lead to the formation of distinct magnetic ground states. However, we have proven that for LiYbO$_2$, a spiral spin liquid ground state is achievable within this new geometry and, therefore, future studies may also probe the robustness of the model to understand the relative importance of the stretching ratio, magnitude of exchange interactions and superexchange pathways when designing new candidate systems.
 
 \textbf{Acknowledgements:} We are grateful to S. D. Wilson (UC Santa Barbara), M. M. Bordelon (Los Alamos National Laboratory) and J. A. M. Paddison (Oak Ridge National Laboratory) for helpful discussions and advice in completing this work. We acknowledge the University of Birmingham and the Institut Laue-Langevin for the PhD Studentship of J. N. Graham and the UKRI Science and Technology Facilities Council for the provision of beamtime.
\bibliography{References}{}
\end{document}